\documentclass[preprint,12pt,authoryear]{elsarticle}

\usepackage{amssymb}
\usepackage{amsthm}
\usepackage{amsmath} 
\usepackage{subfig}
\usepackage{multirow}
\DeclareMathOperator*{\argmax}{argmax}

\journal{Journal of Banking \& Finance}

\begin{document}

\begin{frontmatter}



\title{An Exploration to the Correlation Structure and Clustering of Macroeconomic Variables \tnoteref{t1,t2}}


\author[1]{Garvit Arora\fnref{fn1}}
\ead{garvit.arora@jpmorgan.com}

\author[1]{Shubhangi Tiwari\fnref{fn2}}
\ead{shubhangi20@iitk.ac.in}

\author[3]{Ying Wu\fnref{fn1}}
\ead{ying.wu@jpmchase.com}

\author[3]{Xuan Mei\corref{cor1}\fnref{fn1}}
\ead{xuan.mei@chase.com}

\cortext[cor1]{Corresponding author}

\fntext[fn1]{Garvit, Ying and Xuan are working at the Wholesale Credit QR group}
\fntext[fn2]{Shubhangi is an undergraduate student at Indian Institutes of Technology, Kanpur. She worked at JPMorgan from 2023.06 to 2023.08 as a summer intern}

\affiliation[1]{organization={JPMorgan Chase \& Co.},
            addressline={Level 3 and 4 J.P. Mogan Tower, Off Cst Road Kalina Santacruz East}, 
            city={Mumbai},
            postcode={400098}, 
            country={India}}

\affiliation[3]{organization={JPMorgan Chase \& Co.},
            addressline={545 Washington Blvd.}, 
            city={Jersey City},
            postcode={07310}, 
            state={NJ},
            country={USA}}
\begin{abstract}
As a quantitative characterization of the complicated economy, Macroeconomic Variables (MEVs), including GDP, inflation, unemployment, income, spending, interest rate, etc., are playing a crucial role in banks' portfolio management and stress testing exercise. In recent years, especially during the COVID-19 period and the current high inflation environment, people are frequently talking about the changing "correlation structure" of MEVs. In this paper, we use a principal component based algorithm to perform unsupervised clustering on MEVs so we can quantify and better understand MEVs' correlation structure in any given period. We also demonstrate how this method can be used to visualize historical MEVs pattern changes between 2000 and 2022. Further, we use this method to compare different hypothetical and/or historical macroeconomic scenarios and present our key findings. One of these interesting observations is that, for a list of 132 transformations derived from 44 targeted MEVs that cover 5 different aspects of the U.S. economy (which takes as a subset the 10+ key MEVs published by FRB), compared to benign years where there are typically 20-25 clusters, during the great financial crisis (GFC), i.e., 2007-2010, they exhibited a more synchronized and less diversified pattern of movement, forming roughly 15 clusters. We also see this contrast in the hypothetical CCAR2023 FRB scenarios where the Severely Adverse scenario has 15 clusters and the Baseline scenario has 21 clusters. We provide our interpretation to this observation and hope this research can inspire and benefit researchers from different domains all over the world. 

\end{abstract}





\end{frontmatter}


\section{Introduction}
\label{sec:intro}
As a quantitative characterization of the economy, Macroeconomic Variables (MEVs), such as GDP, inflation, unemployment, income, spending, interest rates, etc., are playing a crucial role in banks' portfolio management and stress testing exercises. For instance, in the CCAR (Comprehensive Capital Analysis and Review, see e.g., \cite{CCAR2020})  and CECL (Current Expected Credit Loss, see e.g., \cite{CECL2016}) exercises, banks develop portfolio and asset-specific models based on historical training data, with MEVs as direct input, and then use these models to obtain loss projections in various hypothetical economic scenarios reflecting experts' outlook of the future US economy. An example is a severely adverse scenario assuming a stagflation and featuring high unemployment and high interest rates simultaneously.  We found \cite{bellini2019ifrs} a good material covering the basics of enlisting MEVs in banking and financial models. 

There are several challenges when considering using MEVs in any models, as we will explain further below. They are connected in certain ways and can be solved to some extent if we understand the correlation structure of MEVs and propose a MEV clustering.

The first difficulty is the high dimensionality and multicollinearity that are intrinsic to MEVs. As one can imagine, to capture the wide spectrum of the economy, a very long list of MEVs is needed. In fact, we see a typical economic scenario scenario for CCAR may consist of 1800 MEVs\footnote{according to Moody's Analytics Economic Scenarios for CCAR https://www.economy.com/products/alternative-scenarios/regulatory-us-federal-reserve-scenarios}, and quickly goes to tens of thousands once we add in simple transformations like quarter-over-quarter (QoQ) and year-over-year (YoY) changes. This immediately brings two difficult and often entangled issues in financial model developmen: high dimensionality of the feature space and the multicollinearity among features. And in many financial problems, due to limited data, these thousands of MEVs and their transformations can even be larger than the total data points (e.g., for some monthly index with only 20 years history there are only 240 data points). There are many statistical tools available to handle these two issues. Ensemble learning methods such as Random Forest \cite{ho1995proceedings} \cite{breiman2001random} and Gradient Boosting Tree \cite{friedman2001greedy} or Stochastic Gradient Boosting Tree \cite{friedman2002stochastic} are amongst statisticians' top picks. In situations where a generalized linear model (aka. GLM, see e.g., \cite{mccullagh2019generalized} Chapter 2) is preferred because of its great transparancy and interpretability, instead of the vanilla Lasso \cite{tibshirani1996regression} or Ridge \cite{hoerl2000ridge}, we can use methods better designed for high dimensionality and multicollinearity, such as Elastic Net \cite{zou2005regularization}, Graphical Lasso \cite{friedman2008sparse} and Group Lasso or Sparse Group Lasso \cite{friedman2010note}. Nevertheless, as Weisberg mentioned in the discussion section of \cite{efron2004least}, with high dimensional multicollinear features, we cannot say for sure, or even trust, that the selected features are the underlying causal variables.  Because of this complication, applying some dimension reduction technique at the very beginning, either guided by domain knowledge or driven by some statistical methodologies, or a combination of these two, can be very helpful. When it comes to statistics based dimension reduction, compared to the supervised method, we prefer an unsupervised approach without any label or target variable. Not only because the former, as discussed by \cite{ambroise2002selection}, when used incorrectly can easily introduce bias and yield spurious results (see \cite{hastie2009elements} section 7.10.2 for a concise example of a common mistake we found pervasive in even published researches), but also in the consideration that an unsupervised method is independent of any target variable and thus the result is more versatile and transferable among models covering different portfolios and asset classes. Considering the prevailance of the GLM in banking and finance space, where the features vector $X$ and its coefficients vector $\beta$ relates to the target variable $Y$ through a linear combination $X\beta$ and a link function $g$, i.e., $g(E(Y|X)) = X\beta$ or $E(Y|X) = g^{-1}(X\beta)$, a correlation based dimension reduction method can be a natural and effective solution.

Secondly, in a model covering a specific portfolio type, among those many MEVs there are some that are more "favorable" than the others, because either they are regarded as fundamental drivers of the economy (e.g., high inflation can lead to a rise of the Federal funds rate and eventually impact most other rates in the market), or are closely related to the portfolio we cover (e.g., Housing Prince Index in MBS and Commercial Price Index in CMBS). As a result, in the model building process, especially those involving parametric models where interpretability and business meaning, as an indispensable layer of defense against overfitting, are often ranked above the pure in-sample fitness, some MEVs are much more preferred to use than others. In general, this preference to a (small) set of MEVs is hard to quantify (i.e., via either weight or penalty) and implement in variable and model selection algorithms. A heuristic, yet simple solution is first using the variable and model selection algorithms (e.g., those we mentioned above) to get a scratch model, and then replacing those selected but less favorable MEVs with their "close neighbors" on the favorable MEVs list. In identifying "close neighbors", a reasonable distance between MEVs is imperative. The correlation structure again is a natural choice.

Lastly, in recently years when the economy was adversely affected by COVID-19, the high inflation and consequently high interest rates, the topic "changing correlation structure of MEVs" is being discussed more frequently within the banking industry, especially the contentious topic of incorporating the COVID-19 period (2020-2022) in modeling. A method to quantify and visualize this change of correlation structure is highly desirable. 

To address these questions, we propose to use a principal component based clustering algorithm mentioned by \cite{anderberg2014cluster} and \cite{sasVarClus}, using the cosine similarity (see e.g., \cite{vipin2006introduction}) that is popular in the data mining and machine learning space. 

The reminder of this paper is structured as follows: Section 2 describes the details of the algorithm that uses principal component analysis and cosine similarity to conduct MEVs clustering. Section 3 lists some interesting applications of this methodology, including an example where we use it to quantify and visualize MEVs correlation structure from 2000 to 2022, and compares different MEVs scenarios. Section 4 concludes this paper.


\section{The Clustering Methodology}
\subsection{Cosine similarity}
\begin{figure}[ht]
	\centering
	\includegraphics[scale=0.4]{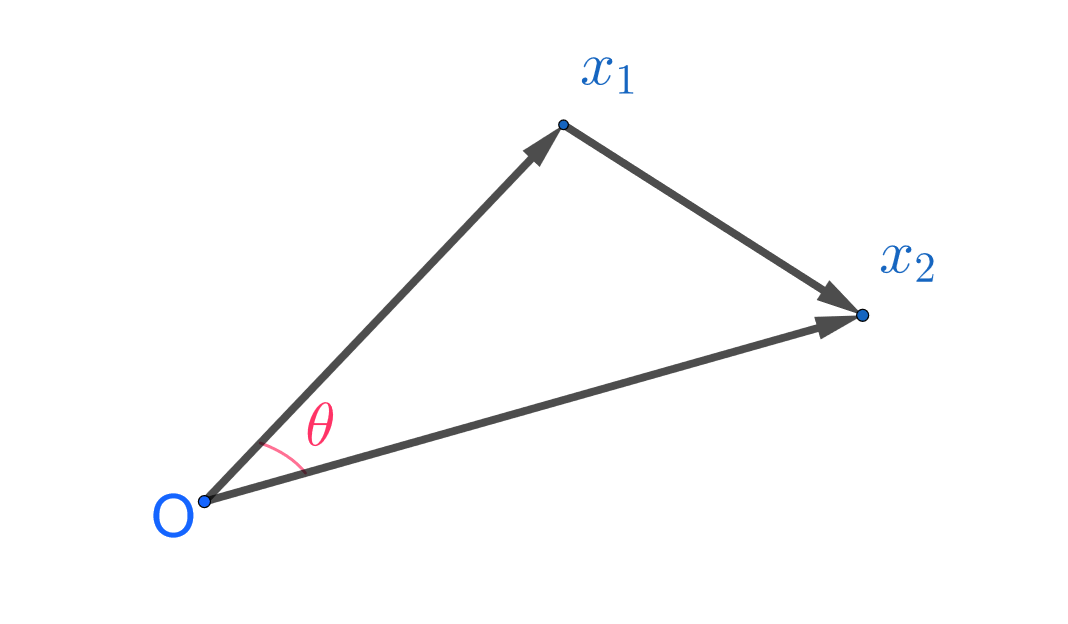}
	\caption{Cosine Similarity for Vectors}\label{fig1}
\end{figure}
Given two real vectors $x_1$ and $x_2$, with the dot product $\langle\cdot, \cdot\rangle$ and Euclidean distance $\Vert \cdot\Vert_2$,  $S_c(x_1, x_2)$, the cosine similarity between $x_1$ and $x_2$, is defined as 
$$ S_c(x_1, x_2) \triangleq \cos(\theta) = \frac{\langle x_1, x_2\rangle}{\Vert x_1 \Vert_2 \cdot \Vert x_2 \Vert_2} $$
This is a quantity that lies between -1 and 1, achieving 1 when these two vectors are proportional (i.e., $\theta = 0$), -1 when they are opposite (i.e., $\theta = \pi$), and 0 when they are perpendicular to each other (i.e., $\theta = \pi/2$).  
 
We prefer cosine similarity to Euclidean distance because the former is closely related to the (Pearson's) correlation. 

It is straightforward that when $x_1$ and $x_2$ are realizations of the random variables $X_1$ and $X_2$ respectively, the cosine similarity between demeaned $x_1$ and $x_2$, i.e., $x_1 - \bar{x}_1$ and  $x_2 - \bar{x}_2$, is simply the sample correlation between $X_1$ and $X_2$, which is a good estimation of the underline true correlation.   

Cosine similarity is also closely related to $R^2$, a.k.a., the coefficient of determination or ratio of variance explained widely used in simple linear regression. If we don't allow intercept, i.e., when we set up $$x_1 = x_2\cdot \beta + \epsilon \textrm{ or } x_2 = x_1\cdot \beta + \epsilon,$$ then we have $$R^2 = S^2_c(x_1, x_2) = \cos^2(\theta)$$ In the more general case intercept is allowed, i.e., $$x_1 = \alpha + x_2\cdot \beta + \epsilon  \textrm{ or }  x_2 = \alpha + x_1\cdot \beta + \epsilon,$$ we have $$R^2 = S^2_c(x_1 - \bar{x}_1, x_2 - \bar{x}_2)$$

\subsection{principal components}
\label{sec:PC}
Principal component analysis (PCA), as a powerful dimension reduction technique, is widely used in the data mining and machine learning space. Detailed introductions are available in almost every data mining and machine learning textbook. We found \cite{jolliffe2016principal} a very good review and summary of this technique and its recent development. 

Let's assume there are  $p$ features $X_1$, $X_2$, $\dots$, $X_p$, and denote as $n$ dimensional column vectors $x_1, x_2, \dots, x_p$ respectively their $n$ observations. Given the fact that in GLM, linear transformations of features (e.g., standardization) doesn't change the ultimate model, we can assume each $x_i$ is centered at 0 (e.g., demeaned). 

For the $n\times p$ dimensional design matrix $X = [x_1, x_2, \dots, x_p]$, its first $k$ ($k<=p$) principal components, each as a $n$ dimensional column vector, can be written as 
\begin{equation}
	PC_i(X) =X\cdot w_i = \sum_{l=1}^{m}w_{il}\cdot x_l, 
\end{equation}
for $i=1, 2, \dots, k$, with the $p$ dimensional column vector $w_i = [w_{i1}, w_{i2}, \dots, w_{im}]^T$ satisfying
\begin{equation*}
	\langle w_i, w_j\rangle = w_i^T \cdot w_j = \sum_{l=1}^{p}w_{i,l}\cdot w_{j,l} =
	\Bigl\{
	\begin{array}{lr}
		1, & \textrm{for } i = j \\
		0, & \textrm{for } i \neq j 
	\end{array}
\end{equation*}

The statistical meaning of principal components is straightforward. Given $p$ random variable $X_1$, $X_2$, $\dots$, $X_p$, we want to find their linear combinations $\sum_{l=1}^{p}w_{i,l}\cdot X_l$ that preserve best their information measured by variance, while being uncorrelated with each other. That is, 
\begin{align*}
w_1 &= \argmax_{\Vert a\Vert = 1}\Bigl\{ \sum_{l=1}^{p}a_l\cdot X_l \Bigl\},\\
w_i &= \argmax_{\Vert a\Vert = 1, a\perp w_j,\forall j<i}\Bigl\{ \sum_{l=1}^{p}a_l\cdot X_l \Bigl\}, i = 2,\dots, k.
\end{align*}
This immediately gives the solution to $w_i$s: By Rayleigh quotient (see e.g., \cite{horn2012matrix}) they are simply eigenvectors corresponding to the top $k$ eigenvalues of the sample covariance matrix of $X$. 

Note that it is straightforward that if $x_i$s are centered (demeaned), then as their linear combinations the principal components are also centered. 

In practice, beside centering, to avoid the case where a feature has large (sample) variance and thus dominates the principal components, we often also rescale the $x_i$s with their sample standard deviation so effectively we deal with standardized $x_i$s and their sample correlation matrix which has all diagonal elements as 1 and thus the trace is $p$. 

The exact method to determine and choose a proper $k$ is not our interest, so we do not elaborate further on this topic. Again, one can refer to various textbooks and papers, such as \cite{vipin2006introduction} and \cite{jolliffe2016principal}.  We want to stress the fact that each principal component is in most cases still a linear combination of all the underline $p$ features, and thus its business and/or economic meaning may sometimes be some what obscure. For example, what does $0.8\cdot\text{Unemployment} -0.6\cdot \text{HPI YoY}$ mean?    

Note that PCA is also related to auto encoder in deep learning. It is equivalent to a single layer auto encoder with linear transfer function. For details, see \cite{plaut2018principal}.

\subsection{Vector clustering using PCA and cosine similarity}
\subsubsection{Split method}
\label{sec:split}
\begin{figure}[ht]%
	\centering
	\subfloat[\centering Calculate the 1st and 2nd PC] {{ \includegraphics[scale=0.15]{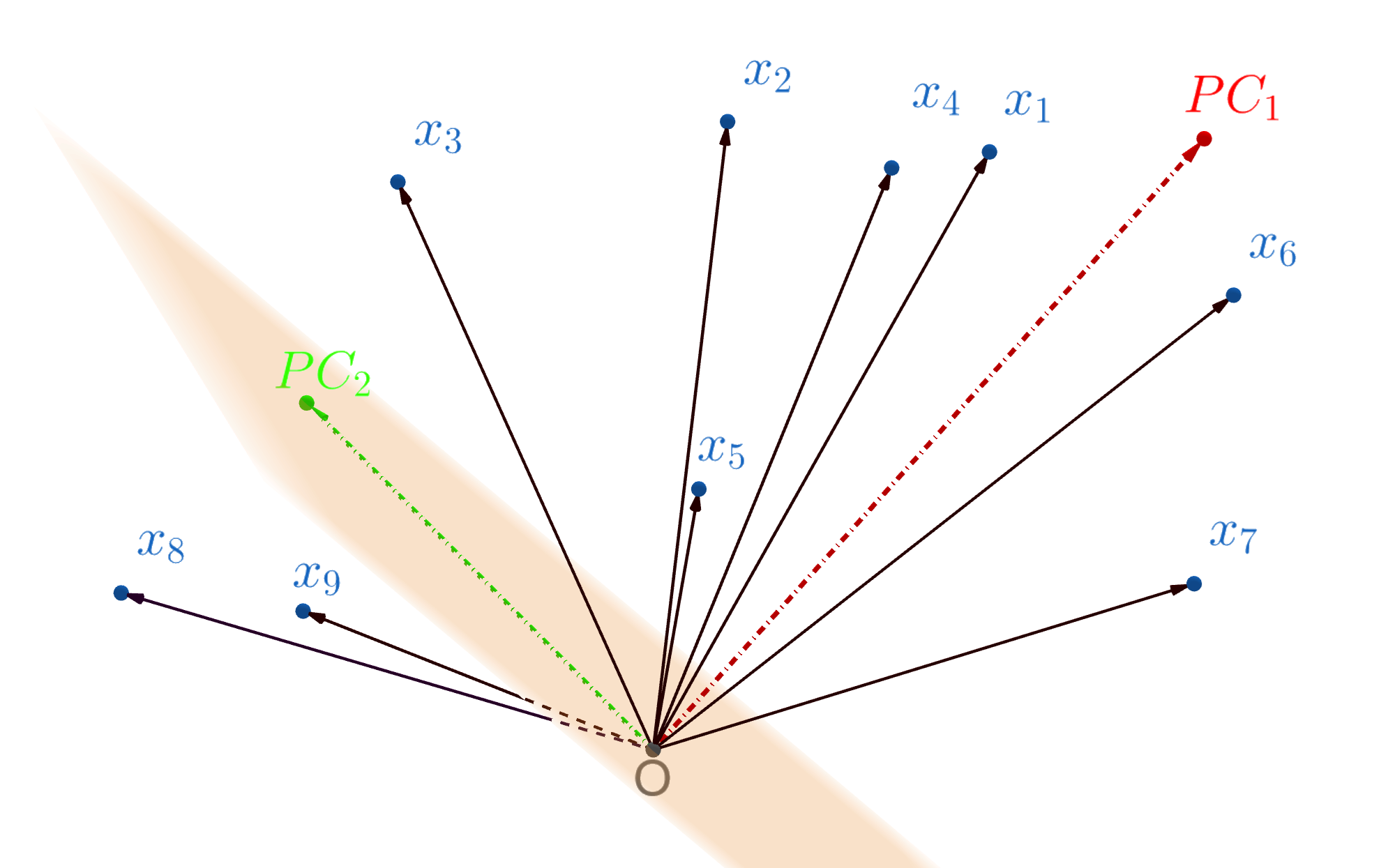} }} %
	\qquad
	\subfloat[\centering Perform Clustering with cosine similarity to 1st and 2nd PC]{{\includegraphics[scale=0.15]{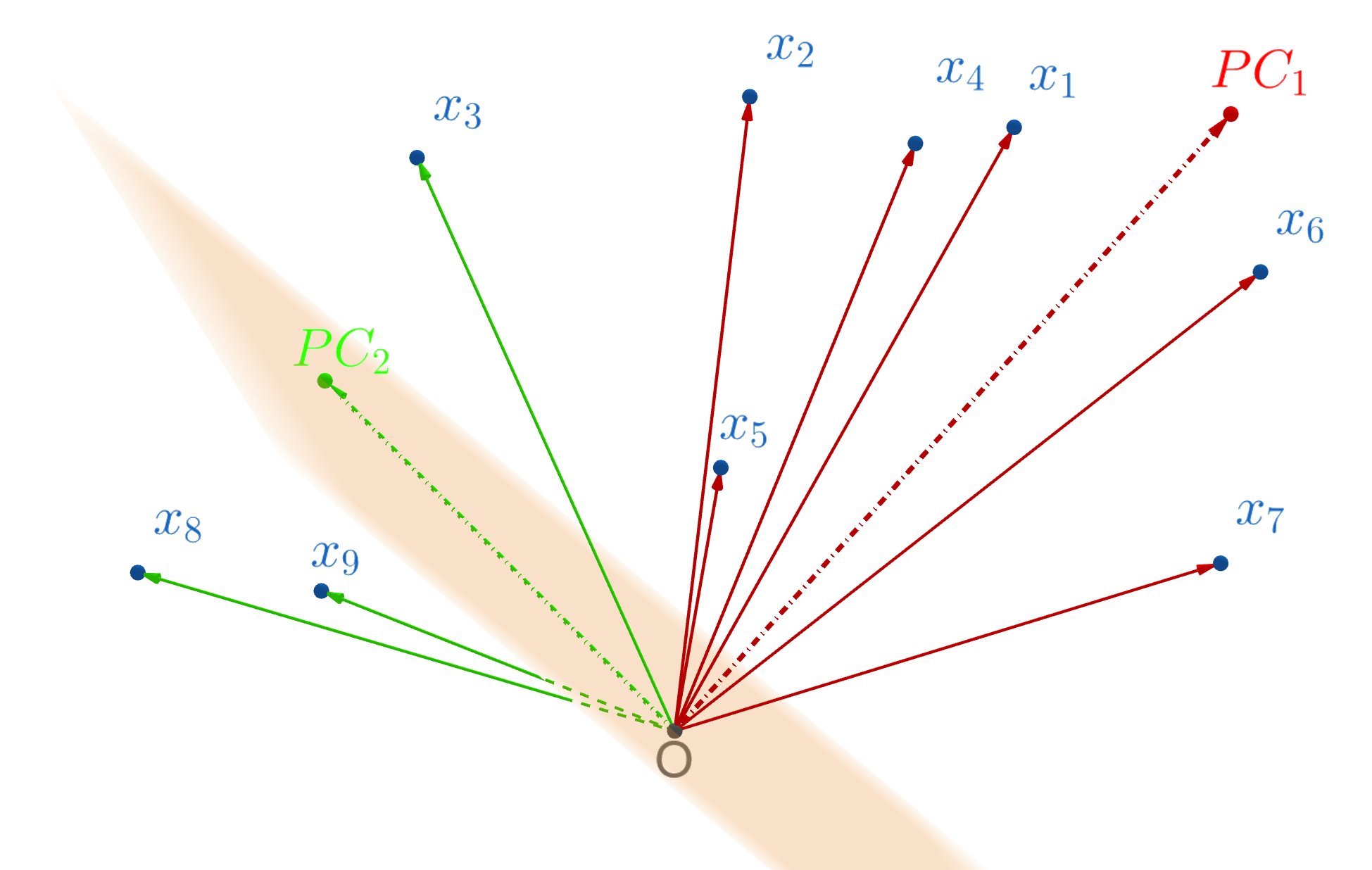} }}%
	\caption{Example of Binary Clustering with principal Components and Cosine Similarity}%
	\label{fig:clustering_step1}%
\end{figure}

\begin{figure}[ht]%
	\centering
	\subfloat[\centering In each new cluster, calculate the new 1st and 2nd PC]{{\includegraphics[scale=0.2]{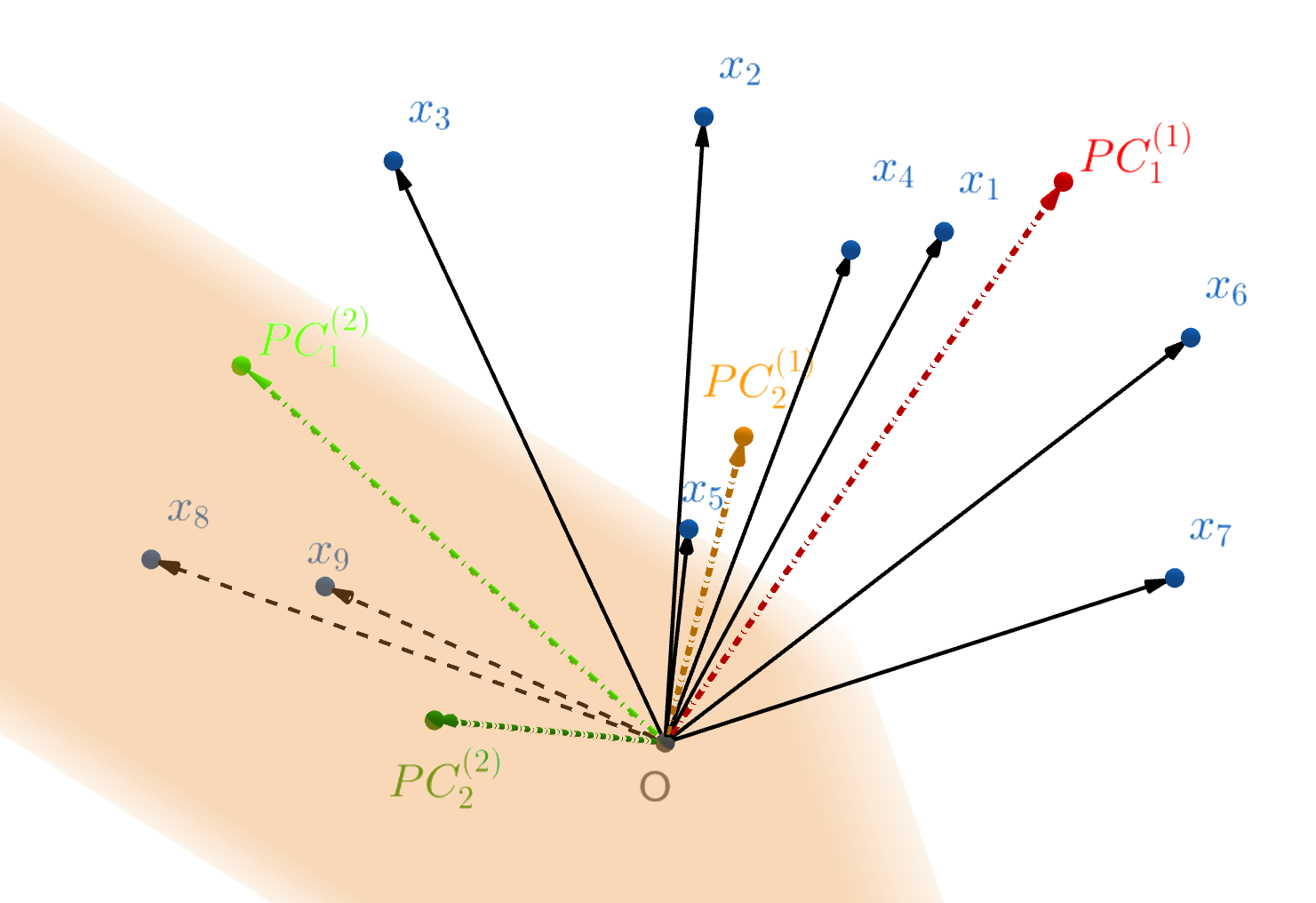} }}%
	\qquad
	\subfloat[\centering In each new cluster, perform clustering with cosine similarity]{{\includegraphics[scale=0.2]{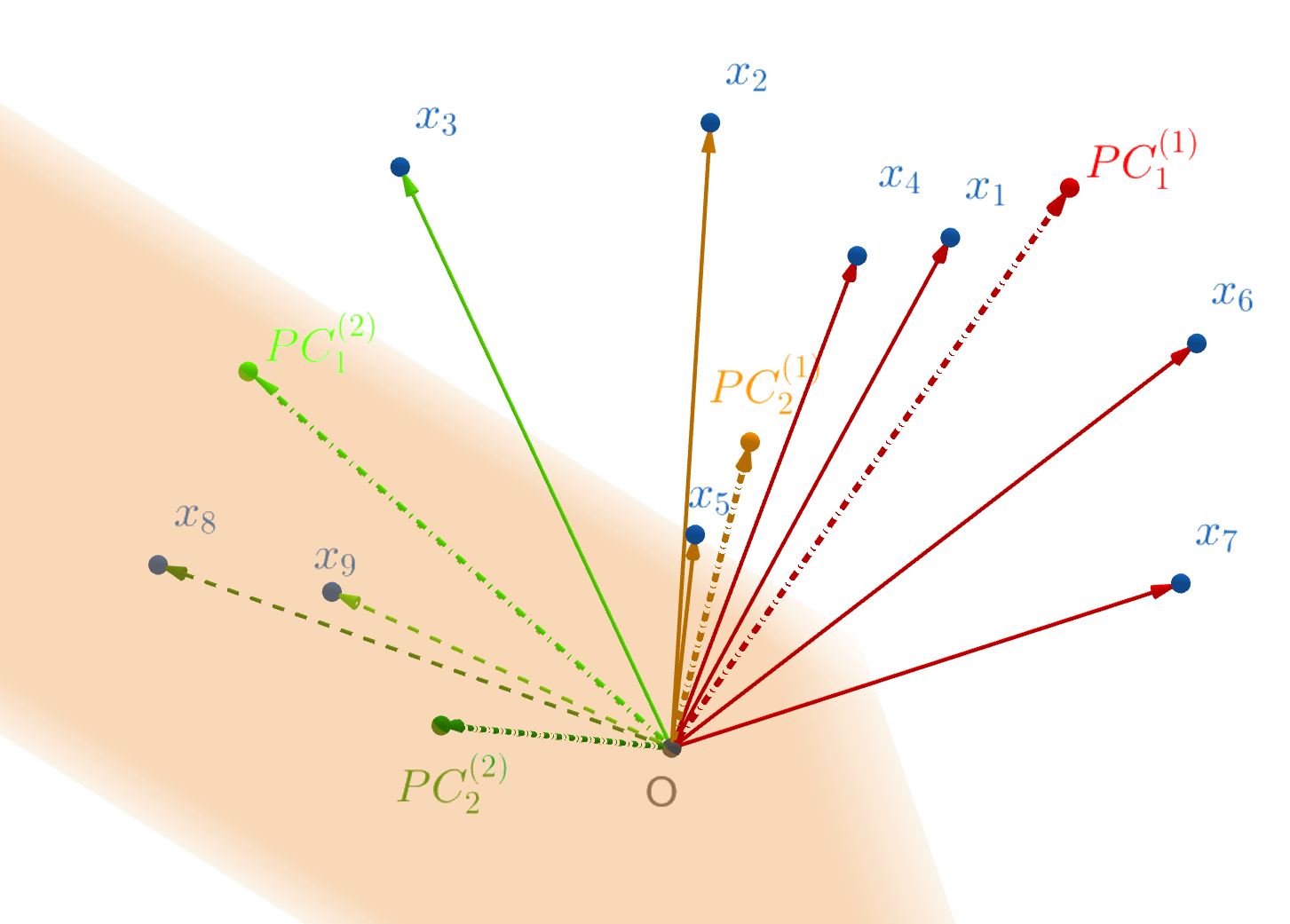} }}%
	\caption{Example of Iterative Binary Clustering }%
	\label{fig:clustering_iterative}%
\end{figure}
For the above $n$ dimensional column vectors $x_1, x_2, \dots, x_p$, we can first calculate their 1st and 2nd principal component $PC_1$ and $PC_2$, which are perpendicular to each other, as shown in Figure \ref{fig:clustering_step1} (a). And then we calculate for each vector $x_i$ its $S_c^2$ (squared cosine similarity) to $PC_1$ and $PC_2$ respectively, and assign it to the cluster formed by the principal component with which it has the higher $S_c^2$, as shown in Figure \ref{fig:clustering_step1} (b). This is very similar to the NCS (nearest centroid sorting) algorithm mentioned by \cite{anderberg2014cluster}. We use $S_c^2$ instead of $S_c$ because only the magnitude of $S_c$ si relevant, not its sign. 

In this way, we split the original whole set $C = \{x_i, i=1, 2, \dots, p\}$ into two disjoint new clusters $C^{(1)}$ and $C^{(2)}$
\begin{align*}
	C^{(1)} &= \bigl\{x\in C: S_c^2(x, PC_1) > S_c^2(x, PC_2) \bigl \}, \\
	C^{(2)} &= \bigl\{x\in C: S_c^2(x, PC_1) \leqslant S_c^2(x, PC_2) \bigl\}.
\end{align*}

We can continue this binary split by first calculating for $C^{(1)}$ its 1st and 2nd principal component, denoted as $PC^{(1)}_1$ and $PC^{(1)}_2$, as shown in Figure \ref{fig:clustering_iterative} (a), and then again split $C^{(1)}$ to $C^{(1,1)}$ and $C^{(1,2)}$, based on the squared cosine similarity, as shown in Figure \ref{fig:clustering_iterative} (b). That is, 
\begin{align*}
	C^{(1,1)} &= \bigl\{x\in C^{(1)}: S_c^2(x, PC^{(1)}_1) > S_c^2(x, PC^{(1)}_2) \bigl \}, \\
	C^{(1,2)} &= \bigl\{x\in C^{(1)}: S_c^2(x, PC^{(1)}_1) \leqslant S_c^2(x, PC^{(1)}_2) \bigl\}.
\end{align*}

Note that, for simplicity we use binary split, similar to the popular binary tree. Multiway,e.g., $k>=3$ way split can be achieved by simply using the first $k$ principal components. 

\subsubsection{Reassignment search}
\label{sec:reassign}
As one may already notice in Figure \ref{fig:clustering_step1} and Figure \ref{fig:clustering_iterative}, $PC_1$ and $PC^{(1)}_1$ are different. This is intuitive since $PC_1$ is the 1st principal component for the whole set while $PC^{(1)}_1$ is on the subset $C^{(1)}$. Also, the 1st principal components in different clusters, e.g., $PC^{(1)}_1$ and $PC^{(2)}_1$, are not necessarily orthogonal to each other. When we use the above iterative splitting algorithm and keep updating the clusters and their principal components, a vector, if assigned to a cluster at certain step, will remain in this cluster and child clusters split from it. However, it is possible that since other clusters and principal components are kept being updated, this vector could be actually closer to principal component from other clusters.

For example, in Figure \ref{fig:clustering_step1} (b) and \ref{fig:clustering_iterative} (b), $x_3$ is first assigned to $C^{(2)}$ and then its child cluster $C^{(2,1)}$. But compared to $PC^{(2)}_1$, it could be closer to $PC^{(1)}_1$ or $PC^{(1)}_2$. Hence in every iteration, after the splitting step we perform a reassignment search step: for every vector we want to calculate its squared cosine similarity to the principal components of other clusters, and reassign it to the cluster it has the maximum squared cosine similarity. And in case a reassignment happens, we need to update the principal components of the two impacted clusters before the next vector is tested. To avoid a local trap where a vector keeps jumping to a new cluster and then back, we don't repeat the search step for this vector and vectors searched before this vector. 

\subsubsection{Stopping criteria}
\label{sec:stopping}
There are a variety of reasonable stopping criteria we can use to stop the above iterative clustering method.

We can use some criteria from tree algorithms. For example, we can use a sparsity based criteria, such as stopping when a cluster has no more than  $m$ (say 5) vectors in it. Or we can consider stopping when vectors in a cluster are already close enough to its 1st principal component, i.e., when in a cluster the minimal or average of squared cosine similarities to its 1st principal component is above a threshold. Or we can simply stop when the maximum number of clusters, e.g., 20, is reached. 

We can also borrow ideas from PCA. For example, we can stop splitting a cluster if its 2nd principal component does not provide much information, i.e., $\lambda_2/\lambda_1$, the 2nd largest eigenvalue divided by the 1st, is below certain threshold. In the most frequent scenario where $x_i$s are already standardized, we know that for a cluster containing $m$ vectors all eigenvalues sum up to $m$ (because the sample covariance matrix has trace $m$, as mentioned in Section \ref{sec:PC}), we can stop when $\lambda_2 < 1$, i.e., when the 2nd principal component provides below average information. We can change this to any number below 1, like 0.8, 0.7, etc.

\subsubsection{Determining the order for splitting}
\label{sec:order}
The above mentioned stopping criteria also provide us a good way to determine which cluster we want to split first when there are a number of clusters available for splitting. 

For example, we can start from the cluster with the largest number of vectors, or the one with the smallest minimum or average similarity, or the one with the largest $\lambda_2/\lambda_1$, or simply the largest $\lambda_2$.

\subsubsection{Summary of the algorithm}
We summarize here the principal component and cosine similarity based clustering algorithm we elaborated above. 

Basically, for a group of vectors we perform the following sketched iterative clustering procedure: 
\begin{enumerate}[Step 1.]
	\item Perform stopping criteria check (see Section \ref{sec:stopping}) and list all clusters that need to split. Stop if the list is empty or the maximum number of clusters is reached.  
	\item Within the above list, determine the first cluster to split (see Section \ref{sec:order})
	\item In this cluster, calculate the 1st and 2nd principal component, and assign each vector to the new cluster formed by the principal component to which it has the higher squared cosine similarity (see Section \ref{sec:split})
	\item Perform a reassignment search on all vectors one by one. If a vector is having higher squared cosine similarity than the principal component of another cluster, reassign it to that cluster and update the principal component of these two clusters. Then go to the next vector (see Section \ref{sec:reassign})

\end{enumerate}
Go to Step 1
 
\section{Clustering on MEVs} 
\label{sec:mev_clustering}
We have been using this clustering methodology intensively in our internal data related to the portfolio we are covering. 

The very basic application is to solve the first two challenges mentioned in Section \ref{sec:intro}. With this clustering methodology, we are able to separate all MEVs and their transformations (after standardization) into clusters. The individual MEVs (transformations) within the same cluster are close to the principal component and thus close to each other in correlation. So in each cluster, we can select one MEV, which exhibits high enough cosine similarity to the principal component while being also on the favorable MEVs list, as the representative of that cluster. Replacing MEVs in the model is also straightforward now. For any MEVs we want to replace, we first identify the MEVs cluster it belongs to, and then replace it using the representative we just mentioned.  

In another example, we used it on the CoStar\footnote{CoStar Group, Inc. is a Washington, DC-based provider of information, analytics, and marketing services to the commercial property industry in the United States, Canada, the United Kingdom, France, Germany, and Spain.} Commercial Real Estate Sales Price Index data that covers roughly 400 US MSA (metropolitan statistical area) and 4 property types (apartment, industry, office and retail), from 2000 to 2022. This is quarterly data so we have $n = 92$ (23 years with 4 quarters per year) and $p = 1600$ (400 MSA and 4 property types). Given the price indices themselves are not stationary we performed 1 month, 3 months, 6 months and 12 months $\log$ difference transformations, respectively. And then we performed clustering at both the overall level (i.e., on all 1,600 vectors) and property type level (i.e., applying clustering only on the 400 vectors for apartment/industry/office/retail, respectively). The clustering result not only tells us which markets have been performing similarly during the past two decades, also by carefully checking the principal components and the "falling apart" markets in each cluster (i.e., those exhibiting low cosine similarity to principal component) we can identify various patterns of this price index and the corresponding markets. More specifically, we can see markets with flat, steadily upward or downward, cyclical, or rather volatile price trend.

Due to the private and proprietary nature of most of our data and projects,  we can only provide some high level introduction of the applications mentioned above and are limited to areas of public research .

As the third and last challenge mentioned in Section \ref{sec:intro}, starting in 2020 conversations about the "changing correlation structure of MEVs" have become more frequent. However, the exact quantification and visualization of either the "correlation structure of MEVs" and "change" are still missing in published papers. Our solution is itemized below. 
\begin{enumerate}
	\item We select a list of target MEVs, denoted as $X$. They are generally those MEVs believed to be relevant to our portfolio. 
	
	For this project we select 44 US national level monthly MEVs from five categories, i.e., general economy, housing market, labor market, rates and spread, including the 16 key MEVs provided by FRB, i.e., Real and Nominal GDP, Real and Nominal Income, unemployment, CPI, 3M/5Y/10Y Treasury Yield, BBB yield, Mortgage rate, Dow Jones Total Stock Market Index, HPI, CRE Price Index and VIX.   
	\item Given the fact that most MEVs are levels and not stationary, we focus on their stationary transformations so the mean and correlation are meaningful and internally consistent. We denote them as $f(X)$.
	
	For this project, we use 3 months, 6 months and 12 months simple differences for all rates like MEVs and 
	$\log$ differences for the rest. So there are eventually $44\times 3$ = 132 MEVs transformations. 
	
	\item For the above $f(X)$, e.g., 132 MEV transformations here, for any interested window of time $T_i$, we apply the MEVs clustering on the realization of $f(X)$ during $T_i$, with the consistent stopping criteria. We denote this clustering operation as $\mathcal{C}(f(X), T_i)$.
	
	In this project, $T_i$s are the annual two-year rolling windows from 2003 to 2022, i.e., $T_1$ is 2003-01 to 2004-12, $T_2$ is 2004-01 to 2005-12, ..., $T_{19}$ is 2021-01 to 2022-12. Thus, $f(X)$ is a 24 (months) by 132 (transformations) matrix for each $T_i$. 
	
    For every two-year rolling window, we first standardize the 132 transformations and then apply clustering. We use a simple stopping criteria $\lambda_2 < 1$ as mentioned in Section \ref{sec:stopping}.
	
	Note that since we have 12 months log or simple differences transformation in $f(X)$, each $T_i$ actually uses three years economic data. We use this setup to sync with the 2-3 years crisis window in both 2008 GFC and various hypothetical stress scenarios. 
	
	\item Utilizing the cardinality operator $|\cdot|$ for sets, we denote as $\mathcal{C}(f(X), T_i)$ the number of clusters in $\mathcal{C}(f(X), T_i)$, and use it as a simple characterization of the target MEVs' correlation structure with window $T_i$. Under the consistent stopping criteria applied to all $T_i$, a bigger $\mathcal{C}(f(X), T_i)$ suggests more diversified underlying MEVs. 	
\end{enumerate} 

\begin{figure}[ht]%
	\centering
	\includegraphics[scale=0.5]{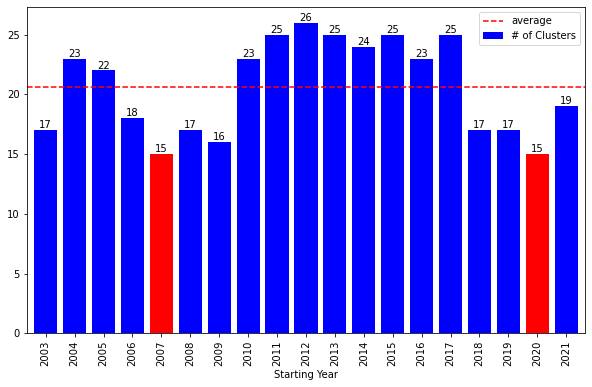} 
    \caption{ Two Years Rolling Window MEVs Clustering Visualization  }%
	\label{fig:MEV_CLustering_Result}%
\end{figure}

In Figure \ref{fig:MEV_CLustering_Result} we visualize such derived number of clusters when clustering those 132 MEV transformations using every two years rolling window. The "Starting Year" axis represents the first/starting year of a two-year window, and the number above the bar is simply the number of clusters. For example, in the two-year window from 2003-01 to 2004-12, the clustering results 17 clusters; and for 2021-01 to 2022-12, it is 19 clusters. For these 19 clusterings, the average number of clusters is 20.6. 

From this visualization it is apparent that we tend to get more clusters in benign years and less clusters in adverse years. The minimum number of clusters, 15, is achieved during the 2007-2008 GFC and the 2020-2021 COVID-19 period. For curiosity, we applied the same clustering on two different hypothetical scenarios coming with CCAR2023 exercise, in the 2-year window from 2023-01 to 2024-12\footnote{In the CCAR2023 exercise, the launch point, i.e., the last actual observation date, is 2022-12-31. MEV curves after this date are hypothetical.}, with the results presented in Table \ref{table:frb}.

\begin{table}[ht]
	\centering
	\begin{tabular}{|l|c|}
		\hline
	\textbf{Hypothetical Scenarios} & \textbf{Number of Clusters} \\ \hline\hline
		FRB Baseline           & 21                 \\ \hline
		FRB Severely Adverse   & 15                 \\ \hline
	\end{tabular}
	\caption{CCAR2023 Hypothetical Scenarios 2-Year Clustering Results}
    \label{table:frb}
\end{table}

Surprisingly, for the FRB Baseline scenario, which represents an averaged outlook for US economy, the consequent number of clusters is 21, very close to the 20.6 average number mentioned above. For the FRB Severely Adverse scenario, we have the exact same 15 clusters as when clustering in 2007-2008 and 2020-2021. Our two internal scenarios, Adverse and Severely Adverse have 14 and 13 clusters, respectively. It therefore still holds that benign years have more clusters compared to severe year.  

One possible explanation to this observation is that, compared to the benign years where the MEVs behave more stably and diversified, during the stress years MEVs tend to move more rapidly in a more synchronized way as the economy crashes, which leads to a stronger correlation structure.  

We further tried CCAR2022 and get relatively similar results: For both Baseline and Severly Adverse scenario, the number of clusters is 2 less than it is in CCAR2023.
\begin{table}[ht]
	\centering
	\begin{tabular}{|l|c|}
		\hline
	\textbf{Hypothetical Scenarios} & \textbf{Number of Clusters} \\ \hline\hline
		FRB Baseline           & 19                 \\ \hline
		FRB Severely Adverse   & 13                \\ \hline
	\end{tabular}
	\caption{CCAR2022 Hypothetical Scenarios 2-Year Clustering Results}
    \label{table:frb}
\end{table}

Our speculation to this observation, i.e., in CCAR2022 both the Baseline and Severely Adverse scenario have 2 less clusters than they have in CCAR2023, is because when FRB was constructing the CCAR2022 scenarios in 2021, the baseline and outlook was rather pessimistic, possibly because of the rampaging covid; while in 2022 everything was stabilized so FRB was less pessimistic when constructing the CCAR2023 scenarios and thus they have 2 more clusters.  

We want to stress the fact that this number of clusters is a simple measure and characterization of the complex correlation structure of MEVs. It can be oversimplified in some cases: Two scenarios with the same number of clusters are not necessarily the same since they can have quite different composition of cluster members. A more thorough approach is to visualize in a graph each cluster and all MEVs in it, so users can see better and conduct more comprehensive comparisons. This is what we have been doing internally. We omit these details here and only present a simple example in order to illustrate the algorithm and its applications.

\section{Summary and Conclusion} 
The macroeconomic variables (MEVs) are widely used in banks' portfolio management and stress testing exercises. While the correlation structure of MEVs is discussed frequently, there are limited publications discussing this topic from a good quantitative perspective. 

In this paper, we aimed to explore the MEV correlation structure by using an unsupervised machine learning algorithm to perform MEV clustering, based on principal components and cosine similarity. We described in detail the underline algorithm, illustrated its geometric meaning and how it is closely related to well-known metrics, such as the Pearson's correlation and $R_2$ in regression. We also demonstrated its usage in dimension reduction, variable selection and replacement, and pattern identification related to MEVs. We gave an example where we used the the MEV clustering to explore and identify historical MEV correlation patterns in a 2-year (24m) rolling window spanning from 2000 to 2022.  We found that for a list of 132 transformations from 44 key MEVs, we observed much less number of clusters in stress years, with 2007-2008 and 2020-2021 achieving the lowest number of clusters (i.e., 15). We apply the same clustering on CCAR2023 hypothetical scenarios and find that in the number of clusters, FRB Baseline and FRB Severely Adverse scenarios align very well with 2000-2022 average and 2007-2008 respectively. Our interpretation to this observation is that, compared to the benign years when the MEVs behave more stably and diversified, during the stress years MEVs tend to move more rapidly in a more synchronized way as the economy crashes, which leads to a stronger correlation structure. 


\newpage

\section*{CRediT authorship contribution statement}
\textbf{Garvit Arora}: Methodology, Resources, Software, Formal analysis, Investigation, Validation,  Data curation, Intern supervision, Writing - review \& editing, Project administration. 
\textbf{Shubhangi Tiwari}: Methodology, Software, Investigation, Formal analysis, Visualization. 
\textbf{Ying Wu}:  Methodology,  Resources. 
\textbf{Xuan Mei}: Conceptualization, Methodology,  Resources, Writing - original draft, Writing - review \& editing.

\section*{Acknowledgements}
We gratefully acknowledge the support and encouragement we received from Rahul Agarwal, Junze Lin and Stuart Marker. We are also deeply indebted to Bjoern Hinrichsen for his heartful suggestions and discussions that drastically improved this paper and made it compliant to JPMC external publication polices. 

\bibliographystyle{elsarticle-harv}
\bibliography{reference}




\end{document}